# Theoretical Study of Temperature Dependence of Spin Susceptibility in Anisotropic Itinerant Ferromagnets

Daisuke Miura* and Akimasa Sakuma

*Department of Applied Physics, Tohoku University, Sendai 980-8579, Japan*

We developed a framework for directly calculating the spin susceptibility of anisotropic itinerant ferromagnets in the full temperature range within the coherent potential approximation in the disordered local moment picture. As a test of our formulation, we demonstrate the computation for the temperature ($T$) dependence of the spin susceptibility for Rashba-type ferromagnets with the Curie temperature $T_\mathrm{C}$. In a certain parameter, we found that the inverse transverse susceptibility $1/\chi_\perp \simeq$ const. for $T < T_\mathrm{C}$ and $1/\chi_\perp \propto T - T_\mathrm{C}$ for $T > T_\mathrm{C}$, which reflects characteristics of itinerant ferromagnets including a spin-orbit interaction.

The magnetic susceptibility $\chi$ is one of the most fundamental properties of magnetic materials, whose microscopic description has attracted the interest of many researchers [1–5]. Historically, demonstrating the Curie–Weiss (CW) law has been a touchstone issue to confirm the reasonableness of models and approximations used in theoretical studies on finite-temperature magnetism. It has long been known that the transverse spin fluctuation plays a central role in the CW behavior in the *localized* electron systems such as ferromagnetic insulators, which can easily be recognized by the Heisenberg model [6]. On the other hand, deriving the CW law on *itinerant* electron systems such as ferromagnetic metals was a lengthy process. Primarily, this is for the itinerant electrons; the magnitude of local moment on each site is not a conserved quantity against the temperature change and the external field, which is governed by the electron-electron correlation effects. For weak ferromagnets, in 1972, Murata and Doniach [7] have first presented the phenomenological description of $\chi$ in which the self-consistent mean field theory was improved to include the mode-mode coupling, and the first microscopic one was given by Moriya and Kawabata in 1973 [8,9]. Furthermore, a unified picture including a treatment of strong ferromagnets was proposed by Moriya and Takahashi in 1978 [10]. In addition to the transverse spin fluctuation considered in the Heisenberg model, these theories considered the longitudinal spin fluctuation through the mode-mode coupling of spin-waves and successfully demonstrated the qualitative characteristic of itinerant electron systems, especially of weak ferromagnets.

Recently, in both practical and industrial viewpoints, strong ferromagnets having a large magnetization and a high $T_\text{C}$ are required. The disordered local moment (DLM) analysis [11–16], by using the first-principles calculation, has shown at quantitative levels that many strong ferromagnetic metals retain robust local magnetic moments even above the $T_\text{C}$, implying that the transverse spin fluctuation greatly governs the CW behavior in these systems. In fact, first-principles calculations for the magnetic susceptibility of Fe [17–20] and Ni [18,20] have been performed on the basis of combined theory [17,21] between the density functional theory and the coherent potential approximation (CPA) in the DLM picture in finite temperatures. There, the CW law was successfully observed for both Fe and Ni, although the calculations for the Curie temperature of Ni resulted in an underestimation indicating the failure of the DLM picture.

In addition to the CW law in $1/\chi$, the importance of finite temperature magnetism of practical materials lies in a temperature dependence of magnetic anisotropy constants (MACs) below $T_\text{C}$. In this temperature range, several authors have performed first-principles calculations for magnetocrystalline anisotropy (MA) in the DLM picture for metallic ferromagnets for FePt [22,23], FePd [23,24], CoPt, MnAl, and FeCo [23]. In addition, the authors demonstrated that the temperature dependence of the MACs are roughly proportional to the second power of the magnetization, as indicated by experiments for FePt [25]. The second power law cannot be understood by single ion models predicting the third power law (so-called Akulov–Zener–Callen–Callen law) [26–31], which suggests a characteristic of ferromagnetic metals that the DLM picture appears to capture well. Again, the transverse susceptibility $\chi_\perp$ can reflect the temperature dependence of MA below the $T_\text{C}$. For example, if a material has uniaxial MA, then the $\chi_\perp$ may have a finite value depending on the temperature, whereas $\chi_\perp$ diverges below the $T_\text{C}$ in the absence of MA. However, theoretical study of $\chi_\perp$ in the full temperature range has not yet been reported for anisotropic itinerant ferromagnetism.

The purpose of this work is to provide a practical framework for *directly* computing the temperature dependence of the magnetic susceptibility of anisotropic ferromagnetic metals in the full temperature range from the microscopic viewpoint based on the DLM and CPA theory. For confirming the validity of our formulation, we also show the temperature dependence of magnetization curves, from whose gradient we can estimate the magnetic susceptibility. However, the computational load for obtaining the magnetization curves is overly heavy; thus, we employ the two-dimensional tight-binding Rashba model [32] for describing itinerant electrons with MA, which has been valued as a theoretically tractable model for investigating anisotropic magnetic metals [33–35]. The usage of the CPA-Green function method enables us to easily extent the present theory to first-principles calculations in the future, which has been widely used for evaluating various physical quantities such as electrical conductivities [36,37], Gilbert damping constants [38,39], magnetic anisotropies [24,40], and so on.

Let us consider the tight-binding Rashba Hamiltonian in the presence of an external uniform static magnetic field $\boldsymbol{H} = (H_x, H_y, H_z)$ on the two-dimensional square lattice given by

$$\mathcal{H} := \sum_{\boldsymbol{k}} c_{\boldsymbol{k}}^\dagger (\hat{\epsilon}_{\boldsymbol{k}} + \hat{v}^H) c_{\boldsymbol{k}} - \Delta_{\text{ex}} \sum_{i=1}^{N} \boldsymbol{e}_i \cdot c_i^\dagger \hat{\boldsymbol{\sigma}} c_i, \tag{1}$$

where $c_{\boldsymbol{k}}$ is the annihilation operator of the electron with a wave vector $\boldsymbol{k} = (k_x, k_y)$ in the spinor representation, the $\boldsymbol{k}$ summation is performed in the first Brillouin zone, $c_i := \left(\frac{1}{\sqrt{N}}\right) \sum_{\boldsymbol{k}} c_{\boldsymbol{k}} \exp(i\boldsymbol{k} \cdot \boldsymbol{R}_i)$, $\hat{\boldsymbol{\sigma}} := (\hat{\sigma}_x, \hat{\sigma}_y, \hat{\sigma}_z)$ for the Pauli matrices $\hat{\sigma}_\alpha$ (a quantity with the hat is a $2 \times 2$ matrix in the spin space), and $\Delta_{\text{ex}}$ denotes the strength of the exchange field whose direction is a classical unit vector $\boldsymbol{e}_i$ allocated at a lattice vector $\boldsymbol{R}_i$ in the $N$ sites. In the language of the functional integral method under the Hubbard-Stratonovich transformation, $\Delta_{\text{ex}}$ corresponds to the saddle point concerning the amplitude of auxiliary exchange fields. The kinetic and Zeeman parts in the first term in Eq. (1), respectively, are defined by

$$\hat{\epsilon}_{\boldsymbol{k}} := -2t(\cos k_x a + \cos k_y a)\hat{1} + \lambda(\hat{\sigma}_x \sin k_y a - \hat{\sigma}_y \sin k_x a), \tag{2}$$

$$\hat{v}^H := \frac{g\mu_B}{2} \hat{\boldsymbol{\sigma}} \cdot \mu_0 \boldsymbol{H} \equiv t\hat{\boldsymbol{\sigma}} \cdot \boldsymbol{h}, \tag{3}$$

where $t$ is the hopping integral between the two nearest-neighbor sites, $a$ is the lattice constant, $\hat{1}$ is the unit matrix, $\lambda$ is the strength of the Rashba spin-orbit interaction (SOI), $g$ is the g factor of an electron, $\mu_B$ is the Bohr magneton, $\mu_0$ is the permeability in vacuum, and $\boldsymbol{h}$ stands for the dimensionless external magnetic field. In the DLM-CPA theory, it is assumed that a local moment at each site is located in a configuration $\{\boldsymbol{e}_i\}$, and $\{\boldsymbol{e}_i\}$ is determined by a functional integral method based on the single-site approximation at each temperature; subsequently, a Green's function is obtained from the CPA condition [11,15]. In the present model, the single-electron Green's function [35] between sites $i$ and $j$ is given by

$$\hat{G}_{ij}^H(z) := \frac{1}{N} \sum_{\boldsymbol{k}} e^{i\boldsymbol{k} \cdot (\boldsymbol{R}_i - \boldsymbol{R}_j)} \hat{g}_{\boldsymbol{k}}^H(z), \tag{4}$$

where $z$ is a complex variable, and $\hat{g}_{\boldsymbol{k}}^H(z) := [z\hat{1} - \hat{\epsilon}_{\boldsymbol{k}} - \hat{v}^H - \hat{\Sigma}^H(z)]^{-1}$. The self-energy $\hat{\Sigma}^H(z)$ is determined by the CPA condition:

$$\langle \hat{t}_e^H(z) \rangle_e^H = \begin{pmatrix} 0 & 0 \\ 0 & 0 \end{pmatrix}, \tag{5}$$

where

$$\langle \cdots \rangle_{\boldsymbol{e}}^H := \int \mathrm{d}\boldsymbol{e}\, w_{\boldsymbol{e}}^H \cdots, \tag{6}$$

$$\int \mathrm{d}\boldsymbol{e} := \int_0^\pi \mathrm{d}\theta \sin\theta \int_0^{2\pi} \mathrm{d}\phi \text{ in terms of } \boldsymbol{e} := (\sin\theta\cos\phi, \sin\theta\sin\phi, \cos\theta), \tag{7}$$

$$\hat{t}_{\boldsymbol{e}}^H(z) := \frac{1}{\hat{1} - [-\Delta_{\mathrm{ex}}\boldsymbol{e}\cdot\hat{\boldsymbol{\sigma}} - \hat{\Sigma}^H(z)]\hat{G}_{00}^H(z)} [-\Delta_{\mathrm{ex}}\boldsymbol{e}\cdot\hat{\boldsymbol{\sigma}} - \hat{\Sigma}^H(z)], \tag{8}$$

$$w_{\boldsymbol{e}}^H := \mathrm{e}^{-\beta(\Omega_{\boldsymbol{e}}^H - F^H)}, \tag{9}$$

$$\Omega_{\boldsymbol{e}}^H := -\frac{1}{\beta}\sum_n \mathrm{Tr}\ln\{\hat{1} - [-\Delta_{\mathrm{ex}}\boldsymbol{e}\cdot\hat{\boldsymbol{\sigma}} - \hat{\Sigma}^H(z_n^H)]\hat{G}_{00}^H(z_n^H)\}, \tag{10}$$

$$F^H := -\frac{1}{\beta}\ln\int \mathrm{d}\boldsymbol{e}\, \mathrm{e}^{-\beta\Omega_{\boldsymbol{e}}^H}, \tag{11}$$

where $\beta := (k_B T)^{-1}$ for the temperature $T$, $z_n^H := i\hbar\omega_n + \mu^H$, $\omega_n := \pi(2n+1)/(\beta\hbar)$, the $n$ summation is performed over all integers, and $\mu^H$ is the chemical potential.

The spin magnetization $\boldsymbol{M}^H$, in units of magnetic field, is represented regarding the single-electron Green's function as

$$\boldsymbol{M}^H = -\frac{g\mu_B}{2a^3\beta}\sum_n \mathrm{Tr}\,\hat{\boldsymbol{\sigma}}\hat{G}_{00}^H(z_n^H) \equiv \frac{g\mu_B}{2a^3}\boldsymbol{\sigma}^h, \tag{12}$$

where $\boldsymbol{\sigma}^h$ denotes the spin concentration vector (dimensionless). Here we assume the form of $\boldsymbol{\sigma}^0 = (0,0,\sigma_s)$ with $\sigma_s > 0$ for $T < T_C$, that is, we set the model parameters such that the system exhibits uniaxially anisotropic ferromagnets. In addition, we must refer to the problem of which be fixed, the chemical potential or the electron concentration. Although the magnetism and the MA of the ground state, at least, must be ferromagnetic and perpendicular to the lattice plane, both strongly depend on the electron concentration as shown by previous theoretical studies on Rashba magnets [33–35]. Therefore, we adjust $\mu^H$ such that the electron concentration is fixed to an appropriate value $n_e$:

$$n_e^H = n_e, \tag{13}$$

where $n_e^H$ is represented by

$$n_e^H = \frac{1}{\beta}\sum_n \mathrm{Tr}\,\hat{G}_{00}^H(z_n^H). \tag{14}$$

From Eq. (12), we obtain the formally microscopic expressions for the longitudinal and transverse spin susceptibilities (dimensionless) as

$$\chi_\parallel := \lim_{H\to 0}\frac{\partial M_z^H}{\partial H_z} = \frac{g^2\mu_B^2\mu_0}{4ta^3}\tilde{\chi}_z, \tag{15}$$

and

$$\chi_\perp := \lim_{H\to 0}\frac{\partial M_x^H}{\partial H_x} = \frac{g^2\mu_B^2\mu_0}{4ta^3}\tilde{\chi}_x, \tag{16}$$

respectively, and the normalized susceptibility has been defined by

$$\tilde{\chi}_\alpha := \lim_{h \to 0} \frac{\partial \sigma_\alpha^h}{\partial h_\alpha} = -\lim_{h \to 0} \frac{1}{\beta} \sum_n \text{Tr}\, \hat{\sigma}_\alpha \frac{\partial \hat{G}_{00}^H(z_n^H)}{\partial h_\alpha}. \tag{17}$$

As indicated, we simply must compute the on-site Green function $\hat{G}_{00}^H(z_n^H)$ in principle, but the high-precision computation is required for reducing an error owing to the numerical differentiation, especially whose computational load for $\chi_\perp$ is heavy. Therefore, developing a method to directly compute the derivative of $\hat{G}_{00}^H(z_n^H)$ is desired for future realistic calculations. It can be achieved by the following procedure.

Firstly, from the definition (4), the derivative of $\hat{G}_{00}^H(z_n^H)$ is represented by

$$\frac{\partial \hat{G}_{00}^H(z_n^H)}{\partial \boldsymbol{h}} = -\frac{1}{N} \sum_{\boldsymbol{k}} \hat{g}_{\boldsymbol{k}}^H(z_n^H) \left( \frac{\partial \mu^H}{\partial \boldsymbol{h}} \hat{1} - t\hat{\boldsymbol{\sigma}} - \frac{\partial \hat{\Sigma}^H(z_n^H)}{\partial \boldsymbol{h}} \right) \hat{g}_{\boldsymbol{k}}^H(z_n^H), \tag{18}$$

and $\partial \mu^H / \partial \boldsymbol{h}$ is determined by

$$\frac{\partial \mu^H}{\partial \boldsymbol{h}} = \left( \frac{1}{\beta N} \sum_{n\boldsymbol{k}} \text{Tr}\, \hat{g}_{\boldsymbol{k}}^H(z_n^H)^2 \right)^{-1} \frac{1}{\beta N} \sum_{n\boldsymbol{k}} \text{Tr}\, \hat{g}_{\boldsymbol{k}}^H(z_n^H)^2 \left( t\hat{\boldsymbol{\sigma}} + \frac{\partial \hat{\Sigma}^H(z_n^H)}{\partial \boldsymbol{h}} \right), \tag{19}$$

which is derived by $\partial n_e^H / \partial \boldsymbol{h} = 0$ from the condition (13) and the definition (14). The final form of the equation to be solved is obtained by differentiating both sides of the CPA condition (5) with respect to $\boldsymbol{h}$:

$$\frac{\partial \hat{\Sigma}^H(z_n^H)}{\partial \boldsymbol{h}} - \left\langle \hat{t}_e^H(z_n^H) \left( \frac{\partial \hat{G}_{00}^H(z_n^H)}{\partial \boldsymbol{h}} - \hat{G}_{00}^H(z_n^H) \frac{\partial \hat{\Sigma}^H(z_n^H)}{\partial \boldsymbol{h}} \hat{G}_{00}^H(z_n^H) \right) \hat{t}_e^H(z_n^H) \right\rangle_e^H$$

$$= -\left\langle \hat{t}_e^H(z_n^H) \text{Tr} \sum_m \left( \frac{\partial \hat{G}_{00}^H(z_m^H)}{\partial \boldsymbol{h}} - \hat{G}_{00}^H(z_m^H) \frac{\partial \hat{\Sigma}^H(z_m^H)}{\partial \boldsymbol{h}} \hat{G}_{00}^H(z_m^H) \right) \hat{t}_e^H(z_m^H) \right\rangle_e^H. \tag{20}$$

Consequently, we can obtain a closed equation for $\partial \hat{\Sigma}^H(z_n^H)/\partial \boldsymbol{h}$ from Eqs. (18), (19), and (20), which is free from the numerical differentiation. After the limit of $\boldsymbol{h} \to \boldsymbol{0}$, the equation for $\partial \hat{\Sigma}^H(z_n^H)/\partial \boldsymbol{h}$ is consisted of only the quantities based on the zero-magnetic field CPA. Here, in Eq. (19) for $T < T_C$ and $\boldsymbol{h} \to \boldsymbol{0}$, we notice that $\partial \mu^H / \partial h_x \to 0$ but $\partial \mu^H / \partial h_z$ does not vanish as noted in Ref. [41].

Let us consider a strong Rashba SOI case [42] in which the parameters are set as $\Delta_{\text{ex}}/t = 0.2$ and $\lambda/t = 1$, and $n_e = 0.25$; the Curie temperature was estimated at $k_B T_C/t = 3.7 \times 10^{-4}$ from the computation of the temperature dependence of $\sigma_s$. Fortunately, because of the simplicity of the present model, we can directly perform the CPA computation for finite $\boldsymbol{h}$. Figure 1 shows the calculated spin-concentration curves along the hard axis, i.e., $\boldsymbol{h} = (h_x, 0, 0)$, by Eq. (12) for several temperatures. For each temperature, we can define a magnetic field $h_A$ such that $\sigma_z^{(h_A,0,0)} = 0$. In the range of $h_x < h_A$, the magnetization rotates in the $z$- $x$ plane as indicated by $|\sigma^h| \simeq$ const. On the other hand, in the range of $h_x \geq h_A$, the magnetization is perfectly parallel to the external magnetic field and the magnitude slowly grows with increasing $h_x$ primarily because of suppressing the spin fluctuation. Here, we should refer the $h_x$ dependence of $\sigma_x^{(h_x,0,0)}$ at a near zero temperature, $T = 0.005 = T_1$ in Fig. 1. Although $\sigma_x^{(h_x,0,0)}$ appears to be in a constant in the high-field range of $h_x > h_A$ in the present scale of the vertical axis in Fig. 1, the gradient — the high-field spin susceptibility $\chi_{high} := \partial \sigma_x^{(h_x,0,0)}/\partial h_x$ — is non-zero, whose value is evaluated at $\chi_{high} \simeq 0.179$ by the linear fitting to $\sigma_x^{(h_x,0,0)}$ with respect to $h_x$. This finite value can be understood as a characteristic of the itinerant electron system. More specifically, $h_x$ enhances the spin polarization of the density of states of the electron via the Zeeman energy, and its effect on $\chi_{high}$ is represented by $\chi_{high} \simeq 4\rho_\uparrow \rho_\downarrow/(\rho_\uparrow + \rho_\downarrow)$ for $h_x \ll \Delta_{ex}/t$ where $\rho_\sigma := -(t/\pi)\Im\left[\hat{G}_{00}^0(\mu^0 + i0)\right]_{\sigma\sigma}$ at $T = 0$. By the expression, we can obtain $\chi_{high} \simeq 0.175$ from the calculated values of $\rho_\uparrow = 0.0909$ and $\rho_\downarrow = 0.0841$, and it is in good agreement with the above fitting value. Next, we can find a high linearity on the $\sigma_x^h$-$h_x$ curves in the cases of $T = T_1, T_2, T_3, T_4 (< T_C)$ and $h_x < h_A$, and can observe that its gradient at $h_x = 0$, i.e., $\tilde{\chi}_x$, does not depend on temperature, in which $\tilde{\chi}_x$ describes the response regarding the magnetization rotation. In the paramagnetic case ($T = T_5, T_6, T_7 (\geq T_C)$), the external magnetic field, even at near zero strength, acts on the magnetization as a suppresser for the spin fluctuation (not as a driving force of the rotation), and thus $\tilde{\chi}_x$ directly reflects the spin fluctuation, and consequently its temperature dependence is led to the CW law. Figure 2 shows the spin susceptibilities calculated on the basis of Eqs. (15)–(20), for $\lambda/t = 1$ and $1/\sqrt{2}$. As expected from the above discussion, $\tilde{\chi}_x$ is a constant for $T < T_C$ and exhibits the CW law for $T \geq T_C$. The comparison between the results for $\lambda/t = 1$ and $1/\sqrt{2}$ suggests $\tilde{\chi}_x^{-1} \propto \lambda^2$, which may be interpreted as below discussion. Now, we approximate $h_A$ by the equation, $\sigma_s = \sigma_x^{(h_A,0,0)} \simeq \tilde{\chi}_x h_A$, that is,

$$h_A = \frac{\sigma_s}{\tilde{\chi}_x}, \tag{21}$$

and if estimating the MAC from $K_u = t\sigma_s h_A/(2a^3)$ by regarding $h_A$ as the magnetic anisotropy field, which corresponds to the area of the triangle formed by $M_z^H$-$\mu_0 H_x$ and $M_x^H$-$\mu_0 H_x$ curves, then we have

$$K_u = \frac{t}{2a^3}\frac{\sigma_s^2}{\tilde{\chi}_x}. \tag{22}$$

Ignoring the effect of the Rashba SOI on the temperature dependence of the saturation magnetization, we may estimate the $\lambda$ dependency of $\tilde{\chi}_x^{-1}$ from one of $K_u$. Because the perturbative treatment for $K_u$ with respect to $\lambda$ gives that $K_u \propto \lambda^2$ [33,34], we can accept the relation $\tilde{\chi}_x^{-1} \propto \lambda^2$. Furthermore, let us consider what Eq. (22) suggests from the viewpoint of the temperature dependence of the MAC. Although the Akulov–Zener–Callen–Callen law predicts $K_\mathrm{u} \propto \sigma_s^3$ for single-site uniaxial MA systems [26–31], several previous studies have indicated $K_\mathrm{u} \propto \sigma_s^2$ in ferromagnetic metals [22–25]. Here, recalling that the present model describes anisotropic ferromagnetic metals, it is reasonable that the temperature dependence of $\tilde{\chi}_\mathrm{x}^{-1}$ becomes a constant, i.e., $K_\mathrm{u} \propto \sigma_s^2$. In the anisotropic Heisenberg model, the temperature dependencies of $\tilde{\chi}_\mathrm{x}^{-1}$ and $\tilde{\chi}_\mathrm{z}^{-1}$ have been demonstrated by several authors [43–45]. Especially, Fröbrich and Kuntz have obtained a constant $\tilde{\chi}_\mathrm{x}$ and a slightly decreasing $\tilde{\chi}_\mathrm{x}$ in inter-site and single-site anisotropic models, respectively, for $T < T_\mathrm{C}$. This suggests that the finite-temperature magnetism of the present model involves the situation presented by the Heisenberg model with the inter-site anisotropy. Lastly, we remark the effect of the SOI on $\tilde{\chi}_\mathrm{z}^{-1}$ in this model. The longitudinal susceptibility is a response for the external magnetic field along the easy axis yielded by the SOI, thus the Zeeman effect and the SOI do not compete. Consequently, the SOI acts as an additional effective magnetic field, only affecting the Curie temperature as indicated by the shift of the zero points of $\tilde{\chi}_\mathrm{z}^{-1}$.

In conclusion, we have developed a computational method for directly calculating the spin susceptibility of itinerant ferromagnets with the uniaxial MA. It is based on the DLM-CPA theory and includes the finite temperature effect over the full range. To demonstrate, we have performed the computation for the temperature dependence of the spin susceptibility in the Rashba-type ferromagnet, and especially, we have confirmed that the transverse susceptibility $\tilde{\chi}_\mathrm{x}$ has a finite value reflecting the MAC for $T \leq T_\mathrm{C}$ and obeys the CW law for $T > T_\mathrm{C}$. For the longitudinal susceptibility $\tilde{\chi}_\mathrm{z}$, on the other hand, the effect of the SOI is limited to the change of $T_\mathrm{C}$ and it shows an ordinary temperature dependence as well as isotropic ferromagnets.

**Acknowledgment**

This work was supported by JSPS KAKENHI Grant Numbers JP17K14800, JP19H05612, and JP21K04624 in Japan.

*E-mail: dmiura1222@gmail.com

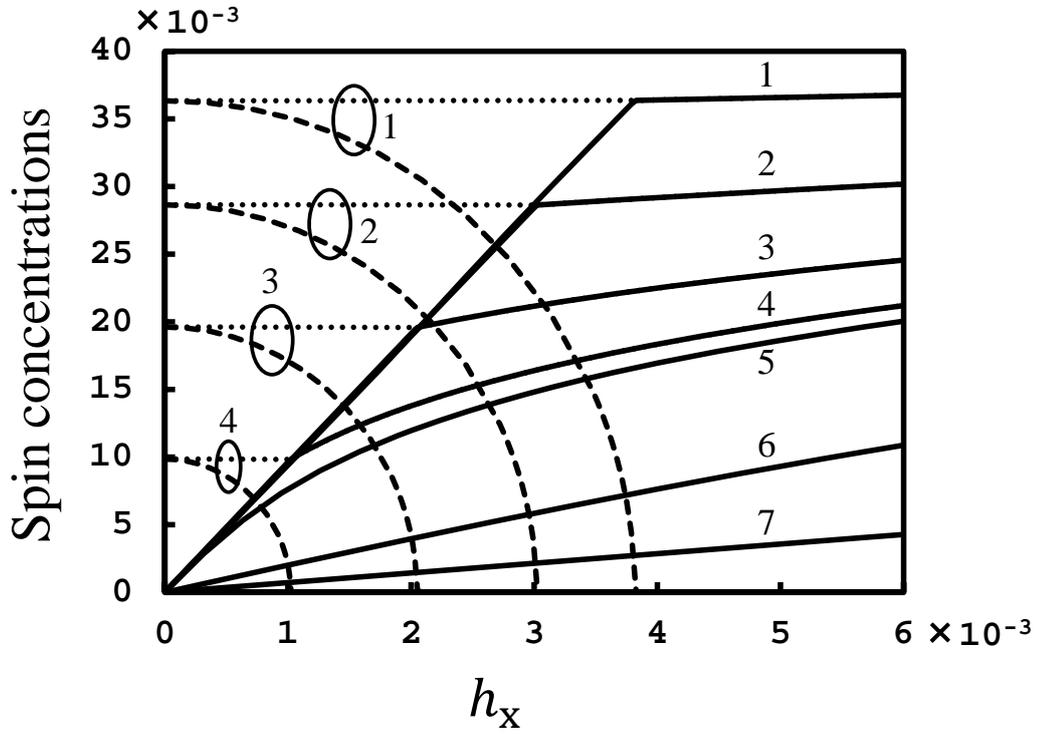

**Fig. 1** The calculated spin concentrations as functions of external magnetic fields $h_x$ for several temperatures. The solid, dashed, and dotted lines are $\sigma_x^h$, $\sigma_z^h$, and $|\boldsymbol{\sigma^h}|$, respectively. Each number reflects the index of the temperatures given by $T_1 = 0.005$, $T_2 = 0.5$, $T_3 = 0.8$, $T_4 = 0.095$, $T_5 = 1$, $T_6 = 1.5$, and $T_7 = 3$ in units of $T_C$. Note that $\sigma_z^h = 0$ and $|\boldsymbol{\sigma^h}| = \sigma_x^h$ for $T_5$, $T_6$, and $T_7$.

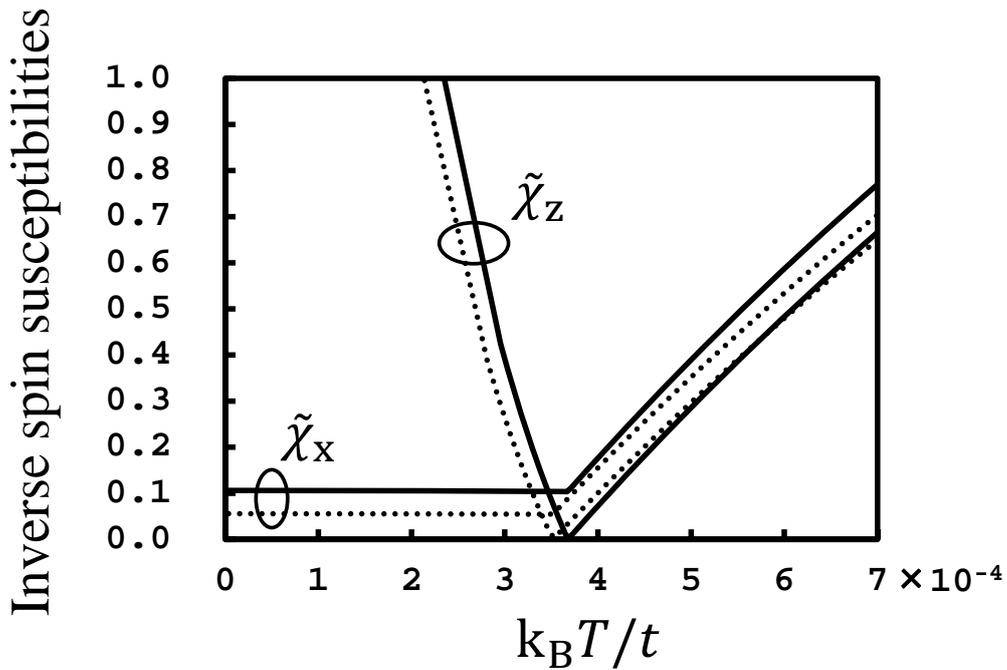

**Fig. 2** The calculated inverse spin susceptibilities as functions of temperature $T$. The solid and dotted lines correspond to the results for $\lambda/t = 1$ and $1/\sqrt{2}$, respectively.